\documentclass[11pt,a4]{article}
\usepackage[top=2.5cm,bottom=2.5cm,left=2.5cm,right=2.5cm]{geometry}
\usepackage{amsmath,latexsym,amssymb,epsfig,psfrag}

\makeatletter   
\@addtoreset{equation}{section}   
\makeatother

\newcommand{\x}{{\bf x}}  
\newcommand{\g}{\Gamma}
\newcommand{\dm}{\textsc{dm}}  
\newcommand{\sm}{\textsc{sm}}  

\newcommand{\rDM}{\rho_\dm}
\newcommand{\drDM}{\dot\rho_\dm}
\newcommand{\rSM}{\rho_\gamma}

\newcommand{\aEM}{\alpha_\textsc{em}}

\newcommand{\mail}[1]{\href{mailto:#1}{{#1}}}
\newcommand{\arxiv}[1]{arXiv:\href{http://arxiv.org/abs/#1}{#1}}
\newcommand{\doi}[2]{\href{https://dx.doi.org/#1}{#2}}
\newcommand{\articoloDOI}[5]{%
  #1, %
  \textit{#2}, %
  \doi{#5}{#3}, \arxiv{#4}.}
\newcommand{\articolo}[3]{%
  #1, \textit{#2}, #3.}
\usepackage[dvipsnames]{xcolor}
\usepackage{hyperref}
\hypersetup{colorlinks, citecolor=ForestGreen, linkcolor=BrickRed, urlcolor=NavyBlue}

\begin{document}  

\begin{titlepage}  
\begin{center}  
\vspace{0.5cm}  
  
\Large {\bf Freeze-in Dark Matter Perturbations are Adiabatic}
  
\vspace*{10mm}  
\normalsize  

{\bf 
D. Racco\footnote{\mail{dracco@stanford.edu}} and A.~Riotto\footnote{\mail{antonio.riotto@unige.ch}}  }

\smallskip   
\medskip   
\it{\small $^1$ Stanford Institute for Theoretical Physics, Stanford University, \\ Stanford, California 94305, USA}

\it{\small $^2$ Department of Theoretical Physics and Gravitational Wave Science Center,  \\
24 quai E. Ansermet, CH-1211 Geneva 4, Switzerland}

\end{center}
  
\vskip0.6in 
   
\centerline{\bf Abstract}
\vskip 1cm  
\noindent
We show that the large-scale perturbations in the dark matter  generated by the freeze-in mechanism are only of  adiabatic nature. 
The freeze-in mechanism is not at odds  with  the  current stringent constraints on isocurvature perturbations.
 
\end{titlepage} 

%%%%%%%%%%%%%%%%%%%%%%%%%%%%%%%%%%%%%%%%%%%%%%%%%%%%%%%%%%%%
%%%%%%%%%%%%%%%%%%%%%%%%%%%%%%%%%%%%%%%%%%%%%%%%%%%%%%%%%%%%
\section{Introduction}  \label{sec:intro}
It is believed that the cosmic microwave background anisotropies and the large-scale structure of the universe are seeded by the fluctuations generated during a period of inflation during the early evolution of the universe \cite{lr}. Such perturbations may be either of adiabatic or of isocurvature nature. Adiabatic, or curvature perturbations, are fluctuations which are generated when there is only
one clock in the universe, e.g. the inflaton field during primordial inflation. Isocurvature perturbations require the presence of more than one degree of freedom and require a quite
subtle dynamics to be produced. 
For instance, isocurvature Dark Matter (DM)  perturbations may be generated by a curvaton field, but only  if the latter decays after the freeze-out temperature of the DM \cite{ung}. 

CMB observations from the Planck collaboration \cite{Planck:2018jri} set stringent constraints on the amount of isocurvature fluctuations present on large-scales. 
A recent study \cite{Bellomo:2022qbx} has claimed that too large isocurvature perturbations in the DM component are generated in the so-called freeze-in production mechanism 
\cite{McDonald:2001vt,Hall:2009bx,Bernal:2017kxu}, when DM particles are generated and never reach chemical equilibrium. 
If true, the freeze-in mechanism would be ruled out  as a production mechanism for the totality of DM. 
However, this conclusion seems at odd with the generic expectation that no isocurvature perturbations may be generated if perturbations may be ascribed to the presence of only one clock \cite{clock,weinberg}. 
In this note, we revisit the issue and indeed conclude that DM perturbations are of adiabatic nature. As we discuss in Sec.~\ref{sec:gauge}, in the freeze-in scenario the energy transferred to DM is only a function of the SM temperature, and the DM pressure is only a function of the DM energy density (and both are only functions of the SM temperature): these two conditions forbid the generation of DM isocurvature perturbations on large scales, because of the absence of another clock. 
In conclusion, the freeze-in mechanism does not suffer of constraints on isocurvature perturbations from the CMB anisotropies. 

\section{Freeze-in DM perturbations: heuristic argument} 
We wish to study the
evolution of cosmological perturbations in the freeze-in transition when DM particles are produced by Standard Model (SM) particle interactions and never reach chemical equilibrium.  

We first offer the reader a heuristic argument of why DM perturbations generated during the freeze-in are adiabatic.
We assume that an adiabatic mode has been produced on superhorizon scales by a period of inflation and communicated to the radiation fluid during the reheating stage \cite{lr}.
We also assume that the SM particles are relativistic and in thermal equilibrium, thus  composing the radiation field. 
Finally, we assume,  for the moment,  that the long mode of the  total curvature perturbation $\zeta_L(\x)$ 
(on length scales $L$ much larger than the Hubble radius $H^{-1}$)  is constant in time. 
One can redefine the local coordinates to absorb the long mode perturbation. For instance, in the uniform energy density gauge, 
the perturbed metric can be written  as 
\begin{equation}
{\rm d}s^2={\rm d}t^2-a^2(t)e^{2\zeta(\x)} {\rm d}\x^2\, ,
\end{equation}
where $a$ is the scale factor. One can absorb the long mode of the curvature perturbation $\zeta_L(\x)$ by simply redefining the spatial coordinates $\x'={\rm exp}(\zeta_L(\x))\x$. In these new coordinates the universe looks homogeneous and isotropic to local observers measuring distances smaller than the Hubble radius.  In other words, we adopt the separate universe approach where it is assumed that each comoving region with size much bigger than the Hubble radius looks locally some unperturbed FRW universe.
 
In the freeze-in mechanism,  the local DM energy density   on scales $\ell \ll H^{-1}$ changes according to the equation (we assume for the moment and for the sake of the argument a  non-relativistic DM)
\begin{equation}
\label{eq:rate}
\dot \rho_\dm(\x',t)+3H \rho_\dm(\x',t)=\g\,,
\end{equation}
where the dot denotes differentiation with respect to the coordinate time
$t$, $\g$ is the production rate (which depends on the number density of the SM particles) and $H$ is the Hubble rate. For instance, 
if DM particles are generated by the
annihilations of SM particles with relative velocity $v_{\rm rel}$ and cross section $\sigma_{\rm ann}$, the production rate reads ($m_\dm$ being the DM masss)
\begin{equation}
\g =m_\dm \langle \sigma_{\rm ann} v_{\rm rel}\rangle n_\sm^2\,.
\end{equation}
To be more concrete, in the freeze-in scenario where the DM is a millicharged fermion $\chi$ with electrical charge $Q_\chi$ produced from pair annihilation of $e^+e^-$, 
\begin{equation}
\g =m_\dm \langle \sigma_\text{ann} v_\text{rel}\rangle  n_e^2 \simeq 
 m_\dm  \left( \frac{9\zeta^2(3)\aEM Q_\chi^2}{2\pi^4}\right) T^4 \,.
\end{equation}
As the SM bath is in thermal equilibrium,  $\g$ is a function only of the local temperature (besides the masses of the SM particles, the DM mass, and   coupling constants), 
\begin{equation}
\g =\g (T(\x',t))\, ,
\end{equation}
and therefore the local DM  number density after the freeze-in  is  only a function of the local temperature $T({\bf x'},t)$.
The latter  inherits the   large-scale superhorizon fluctuations seeded by inflation \cite{np}
\begin{equation}
T(\x',t)=T_{\rm bg}(t)e^{\zeta_L(\x)/5}\, ,
\end{equation}
where $T_{\rm bg}$ is the background temperature. 
This already suggests that  there is only one clock governing the fluctuations of the freeze-in DM, the one generated by the curvature perturbation $\zeta_L(\x)$, and that no isocurvature perturbation can be generated in the absence of an additional separate source of perturbations \cite{weinberg}. 
As we will see, the gauge-invariant expression for the DM-radiation isocurvature perturbation  depends only  on the difference
\begin{equation}
\frac{\delta \rDM(\x,t)}{ \drDM}- \frac{\delta \rSM(\x,t) }{\dot\rho_\gamma}\, .
\end{equation}
Since
\begin{equation}
\frac{\delta \rDM(\x,t)}{ \drDM}=\frac{\delta T(\x,t)}{\dot T} = \frac{\delta \rSM(\x,t) }{\dot{\rho}_\gamma}\, ,
\end{equation}
we expect that no isocurvature perturbations are generated through the freeze-in mechanism in the DM component, in agreement as well with our previous assumption that the total curvature perturbation is constant in time on superhorizon scales.
In the following we will arrive at the same conclusion through a rigorous gauge-invariant treatment.

\section{Freeze-in DM perturbations: the gauge-invariant treatment } 
To study the freeze-in DM perturbations in a gauge-invariant manner, 
we follow the gauge-invariant approach developed in Ref.~\cite{wmu}
for the general case of an arbitrary number of interacting fluids in general relativity. 

\subsection{The background equations}
The evolution of the background FRW universe during the freeze-in stage is governed by the Friedmann constraint
\begin{align}
\label{Friedmann}
H^2 &= \frac{8\pi G}{3}\rho \,,\\
\dot H &= -4\pi G \left( \rho+P\right)
 \,,
\end{align}
and the continuity equation
\begin{equation}
\label{continuity}
\dot\rho=-3H\left( \rho+P\right)\,,
\end{equation}
where
$\rho$ and $P$ are the total energy density and the total
pressure of the system.
The total energy density
and the total pressure are related to the energy density and
pressure of the DM field and radiation  by
\begin{align}
\rho &=\rho_\dm+\rho_\gamma \,, \nonumber\\
P&= P_\dm+P_\gamma \,, 
\end{align}
where $P_\gamma$ is the radiation pressure. 
The DM field and the radiation component 
have energy-momentum tensor $T^{\mu\nu}_{\dm}$ and 
$T^{\mu\nu}_{\gamma}$, respectively.
The total energy momentum tensor 
\begin{equation}
T^{\mu\nu}=T^{\mu\nu}_{\dm}+
T^{\mu\nu}_{\gamma}
\end{equation}
is covariantly conserved, but we allow for
energy transfer between the fluids,
\begin{align}
 \label{Qvector}
\nabla_\mu T^{\mu\nu}_{\dm}&=Q^\nu_{\dm}\,,\nonumber\\
\nabla_\mu T^{\mu\nu}_{\gamma}&=Q^\nu_{\gamma}\,, 
\end{align}
where $Q^\nu_{\dm}$ and $Q^\nu_{\gamma}$ are
 the generic energy-momentum transfer to
the inflaton and radiation sector respectively
and are  subject to the constraint
\begin{equation}
\label{Qconstraint}
Q^\nu_{\dm}+Q^\nu_{\gamma}=0 \,.
\end{equation}
The continuity equations for
the energy density of the DM field  and radiation 
 in the background are thus ($Q_\dm=Q_{\dm}^0$, 
$Q_\gamma=Q_{\gamma}^0$)
\begin{align}
\label{m}
\dot\rho_{\dm}
&=-3H\left(\rho_{\dm}+P_{\dm}\right) +Q_{\dm}\,,\\
\dot\rho_{\gamma}
&=-3H\left(\rho_{\gamma}+P_{\gamma}\right) +Q_{\gamma}\,.\nonumber
\end{align}
In the following  we parametrise the 
energy transfer between radiation and the DM by 
\begin{gather}
\label{defbackQa}
Q_{\dm} = \g(\rho_{\gamma}) \,, \nonumber\\
\label{defbackQb}
Q_{\gamma} =- \g(\rho_{\gamma})\, .
\end{gather}
This assumption is motivated  again by the fact that in the freeze-in mechanism DM particles are generated out of chemical equilibrium and by SM degrees of freedom
which are relativistic and therefore a function of only the temperature or radiation energy density (besides the masses and coupling constants).
The background energy conservation equations therefore read
\begin{align}
\label{dotrhos}
\dot\rho_\dm &= -3H(\rho_\dm+P_\dm)+\g\,, \\
\label{dotrhog}
\dot\rho_\gamma &= -4H\rho_\gamma-\g\,, \\
\label{dotrhotot}
\dot\rho&=-H\big[3(\rho_\dm+P_\dm)+4\rho_\gamma\big] \,.
 \end{align}

%%%%%%%%%%%%%%%%%%%%%%%%%%%%%%%%%%%%%%%%%%
\subsection{Gauge-invariant linear perturbations}\label{sec:gauge}
%%%%%%%%%%%%%%%%%%%%%%%%%%%%%%%%%%%%%%%%%%

Linear scalar perturbations about a
spatially-flat FRW background model are defined by the line
element \cite{lr}
\begin{equation} 
{\rm d}s^2=-(1+2\varphi){\rm d}t^2+2aB_{,i}{\rm d}t {\rm d}x^i
+a^2\left[(1-2\psi)\delta_{ij}+2E_{,ij}\right]{\rm d}x^i{\rm d}x^j \,, 
\end{equation}
where we have introduced the 
gauge-dependent curvature perturbation, $\psi$, the lapse
function, $\varphi$, and scalar shear, $\chi\equiv a^2\dot E - aB$.
The perturbed energy transfer rates
including terms up to first order, are  written as
\begin{equation}
  - Q_{\dm}(1+\varphi) - \delta Q_\dm\,\,\,{\rm and}\,\,\,
 -  Q_{\gamma}(1+\varphi) - \delta Q_\gamma\,,
\end{equation}
where the gravitational redshift (time-dilation) factor $(1+\varphi)$
has been made manifest.
Both the density perturbations  $\delta\rho_\dm$ and $
\delta\rho_\gamma$ and the
gravitational potential  $\psi$ are in general gauge-dependent. 
However  gauge-invariant combinations can
be constructed which describe the density perturbations on uniform
curvature slices or, equivalently the curvatures of uniform density
slices. The total curvature perturbation $\zeta$ on uniform total density hypersurfaces is given by 
\begin{equation}
\label{zeta}
\zeta=-\psi-H\frac{\delta\rho}{\dot\rho} \,,
\end{equation}
while the curvature perturbation on uniform DM energy density
and radiation energy density
hypersurfaces are respectively  defined as
\begin{align}
\label{zetaalpha}
\zeta_{\dm}&=-\psi-H\frac{\delta\rho_\dm}{\dot{\rho}_\dm} \,,
\nonumber\\
\zeta_{\gamma}&=-\psi-H\frac{\delta\rho_\gamma}{\dot{\rho}_\gamma} \,.
\end{align}
The total curvature perturbation (\ref{zeta}) is thus a weighted
sum of the individual perturbations
\begin{equation}
\label{zetatot}
\zeta= 
 \frac{\dot{\rho}_\dm}{\dot\rho}  \zeta_\dm+
\frac{\dot{\rho}_\gamma}{\dot\rho}  \zeta_\gamma \,,
\end{equation}
while the difference between the two curvature perturbations
describes a relative gauge-invariant entropy (or isocurvature) perturbation%
\begin{equation}
 \label{defS}
  {\cal S}_{\dm\gamma}=3(\zeta_\dm-\zeta_\gamma)
   = -3H
\left(
  \frac{\delta\rho_\dm}{\dot{\rho}_\dm}
  - \frac{\delta\rho_\gamma}{\dot{\rho}_\gamma} \right) \, .
\end{equation}
From the definitions of the total curvature perturbation
(\ref{zetatot}) and the entropy perturbation (\ref{defS}), we get
for instance that 
\begin{equation}
\begin{aligned}
 \label{relation}
 \zeta_{\dm}&=\zeta+\frac{1}{3} 
  \frac{\dot{\rho_\gamma}}{\dot\rho}{\cal S}_{\dm\gamma}\\
  \zeta_{\gamma}&=\zeta-\frac{1}{3} 
  \frac{\dot{\rho}_\dm}{\dot\rho}{\cal S}_{\dm\gamma}\,.
\end{aligned}
\end{equation}
On wavelengths larger than the horizon scale, 
the perturbed energy conservation equations  for the DM energy
density and the radiation energy density can be written, 
including energy transfer, as
\begin{align} \label{pertenergyexact}
\dot{\delta}\rho_{\dm}+3H(\delta\rho_{\dm}+\delta P_{\dm})
- \left(\rho_{\dm}+P_{\dm}\right)3\dot\psi
&=  Q_{\dm}\varphi+\delta Q_{\dm}\,,\nonumber\\
\dot{\delta}\rho_{\gamma}+3H(\delta\rho_{\gamma}+\delta P_{\gamma})
- \left(\rho_{\gamma}+P_{\gamma}\right)3\dot\psi
&=   Q_{\gamma}\varphi+\delta Q_{\gamma}\,.
\end{align}
Using the perturbed $(0i)$-component
of Einstein's equations  for super-horizon wavelengths
\begin{equation}
\dot\psi+H\varphi=-\frac{H}{2}\frac{\delta\rho}{\rho}\,, 
\end{equation}
we can 
re-write Eq. (\ref{pertenergyexact})
  in terms of the gauge-invariant curvature
perturbations $\zeta_\dm$ and $\zeta_\gamma$   \cite{wmu}
\begin{align}
 \label{dotzetaalpha}
\dot\zeta_\dm &=
 -\frac{H}{\dot{\rho}_\dm}\left(\delta Q_{\rm{intr},\dm}+
\delta Q_{\rm{rel},\dm}\right) + \frac{3H^2}{\dot\rho_\dm}\delta P_\text{intr,\dm}
 \,,\nonumber\\
\dot\zeta_\gamma &=-\frac{H}{\dot{\rho}_\gamma}
 \left(\delta Q_{\rm{intr},\gamma}+
\delta Q_{\rm{rel},\gamma}\right) + \frac{3H^2}{\dot\rho_\gamma}\delta P_{\text{intr},\gamma}\,,
\end{align}
where
\begin{equation}
\label{deltaQintralpha}
\begin{aligned}
\delta Q_{{\rm intr},\dm} &\equiv \delta Q_\dm -
{\dot{Q}_\dm\over\dot{\rho}_\dm} \delta\rho_\dm \,, 
& \quad \delta P_{{\rm intr},\dm} &\equiv \delta P_\dm -
{\dot{P}_\dm\over\dot{\rho}_\dm} \delta\rho_\dm \,,
\\
\delta Q_{{\rm intr},\gamma} &\equiv \delta Q_\gamma -
{\dot{Q}_\gamma\over\dot{\rho}_\gamma} \delta\rho_\gamma\,,
& \delta P_{{\rm intr},\gamma} &\equiv \delta P_\gamma -
{\dot{P}_\gamma\over\dot{\rho}_\gamma} \delta\rho_\gamma\,,
\end{aligned}
\end{equation}
are the gauge-invariant perturbations for the intrinsic non-adiabatic energy transfer and the pressure, and
\begin{align}
\label{deltaQrelalpha}
\delta Q_{{\rm rel},\dm} &=
{  Q_\dm\dot\rho \over 2\rho} \left(
{\delta\rho_\dm\over\dot{\rho}_\dm} - {\delta\rho\over\dot\rho}
\right)
= - {  Q_\dm\over 6H\rho}  \dot{\rho}_\gamma {\cal S}_{\dm\gamma}\,,\nonumber\\
\delta Q_{{\rm rel},\gamma} &=
{  Q_\gamma\dot\rho \over 2\rho} \left(
{\delta\rho_\gamma\over\dot\rho_\gamma} - {\delta\rho\over\dot\rho}
\right)
= +{  Q_\gamma \over 6H\rho}  \dot\rho_\dm {\cal S}_{\dm\gamma}
\end{align}
are the relative  gauge-invariant non-adiabatic perturbed energy transfer 
due to the presence of relative entropy perturbations \cite{wmu}.
The intrinsic pressure perturbations $\delta P_\text{intr}$ for radiation and DM both vanish. Indeed, for the radiation one has $P_\gamma=\rho_\gamma/3$ and for  the DM, even for a non-thermal DM phase space density, one has that both $P_\dm$ and $\rho_\dm$ are functions of the thermal temperature and therefore one can always write the pressure $P_\dm$ as a function of the energy density $\rho_\dm$ (in the relativistic limit one would have 
 $P_\dm=\rho_\dm/3$, which  valid even for a non-thermal  DM distribution in phase space; in the non-relativistic limit  $P_\dm=0$).

The evolution equations (\ref{dotzetaalpha}) for the
curvature perturbation on uniform DM  density hypersurfaces,
$\zeta_\dm$, and uniform radiation density hypersurfaces,
$\zeta_\gamma$, are given by
\begin{align}
\label{dotzetacurv}
\dot\zeta_\dm
&=
{ \g \over6\rho} 
{\dot\rho_\gamma\over\dot\rho_\dm}{\cal S}_{\dm\gamma}-\frac{H}{\dot\rho_\dm}\delta\g_\dm
 \,, \\
\label{dotzetarad}
\dot\zeta_\gamma
&={ \g \over6\rho} 
{\dot\rho_\dm\over\dot\rho_\gamma}{\cal S}_{\dm\gamma}-\frac{H}{\dot\rho_\gamma}\delta\g_\gamma\,, 
\end{align}
where \cite{mr}
\begin{align}
\delta\g_\dm&=\delta\g-\dot\g
\frac{\delta\rho_\dm}{\dot\rho_{\dm}}\,,\nonumber\\
\delta\g_\gamma&=-\delta\g+\dot\g
\frac{\delta\rho_\gamma}{\dot\rho_{\gamma}}
\end{align}
are the gauge-invariant perturbations of the freeze-in production rate. Taking $\g=\g(\rho_\gamma)$, we find
\begin{align}
\delta\g_\dm&=\delta\g-\dot\g
\frac{\delta\rho_\dm}{\dot\rho_{\dm}}=\dot\g\left(\frac{\delta\rho_\gamma}{\dot\rho_{\gamma}}-\frac{\delta\rho_\dm}{\dot\rho_{\dm}}\right)=\frac{\dot\g}{3H}{\cal S}_{\dm\gamma}\,,\nonumber\\
\delta\g_\gamma&=0\, .
\end{align}
We then obtain
\begin{align}
\label{dotzetacurv1}
\dot\zeta_\dm
&=\left(
{ \g \over6\rho} 
{\dot\rho_\gamma\over\dot\rho_\dm}-\frac{\dot\g}{3\dot\rho_\dm}\right){\cal S}_{\dm\gamma}=\frac{\dot\rho}{\dot\rho_\dm}
\left({ \g \over2\rho} 
{\dot\rho_\gamma\over\dot\rho_\dm}-\frac{\dot\g}{\dot\rho_\dm}\right)(\zeta-\zeta_\gamma)
 \,, \\
\label{dotzetarad1}
\dot\zeta_\gamma
&={ \g \over6\rho} 
{\dot\rho_\dm\over\dot\rho_\gamma}{\cal S}_{\dm\gamma}=\frac{\dot\rho}{\dot\rho_\gamma} {\g \over2\rho} 
{\dot\rho_\dm\over\dot\rho_\gamma}(\zeta_\dm-\zeta)=\frac{\dot\rho}{\dot\rho_\gamma} {\g \over\rho} 
(\zeta-\zeta_\gamma)\,. 
\end{align}
It follows that
\begin{equation}
\dot{\cal S}_{\dm,\gamma} = \frac{3\dot\rho}{\dot\rho_\dm^2}
  \left( \frac{\dot \rho_\gamma^2 - \dot \rho_\dm^2}{\dot\rho_\gamma} \frac{\g}{2\rho} - \dot\g \right) (\zeta-\zeta_\gamma)\, .
\end{equation}
Now, the initial condition of  such evolution equations, before the freeze-in transition, is
\begin{equation}
\zeta(\x,t\ll t_{\rm fi})=\zeta_\gamma(\x,t\ll t_{\rm fi}),
\end{equation}
from which we conclude that 
\begin{equation}
\zeta(\x,t\gg t_{\rm fi})=\zeta_\gamma(\x,t\gg t_{\rm fi})\,\,\,{\rm and}\,\,\, {\cal S}_{\dm\gamma}(\x,t\gg t_{\rm fi})=0
\end{equation}
are fixed points of the evolution. DM perturbations inherit the same adiabatic perturbations generated during inflation. 
This rigorous result confirms the argument that no isocurvature perturbations may be generated in the presence of only one clock. 

Let us conclude with the following comment. Our conclusion is valid also for the DM produced out of chemical equilibrium through a decay of a SM degree of freedom 
whose perturbation is adiabatic, that is of the curvature type. In such a case, one will have again simply $\zeta_\dm=\zeta$. 
In general, the situation with the production of the DM by the freeze-in mechanism through interactions of the SM particles which dominate the energy density of the universe and whose perturbations are curvature perturbations is completely analogous to what happens in the case in which DM particles are produced 
by the curvaton decay if the latter happens after the freeze-out epoch. Such DM particles do not  reach chemical equilibrium and if the curvaton at the time of its decay dominates the energy density, it simply transfers its curvature perturbation to the DM, with no residual isocurvature perturbation  \cite{ung}.

As a final remark, we notice that the definition of the isocurvature perturbation in Eq.~\eqref{defS} is gauge-invariant also if there is an energy transfer between radiation and DM (as it is the case for freeze-in production). 
This is not the case for the alternative definition of Kodama and Sasaki \cite[Eq.~(5.38)]{Kodama:1984ziu}, in presence of energy transfer: this can be easily seen by writing that expression in a gauge where $\delta T^{0i}=0$, so that 
\begin{equation}
{\cal S}^\text{[KS]}_{\dm\gamma} = 
\frac{\delta \rho_\dm/\rho_\dm}{1+w_\dm}-\frac{\delta \rho_\gamma/\rho_\gamma}{1+w_\gamma}=
-3H\left(\frac{\delta \rho_\dm}{\dot\rho_\dm-Q_\dm}-\frac{\delta \rho_\gamma}{\dot\rho_\gamma-Q_\gamma}\right)
\end{equation}
whose evolution equations \cite{Kodama:1984ziu, Hamazaki:1996ir} are more involved than those in Ref.~\cite{wmu}, that we discuss here.
Maybe this could be the source of the disagreement with Ref.~\cite{Bellomo:2022qbx}.

\vskip 0.5cm
\centerline{\bf Acknowledgements}
\vskip 0.2cm
\noindent
D.R.\ thanks Sebastian Baum and the phenomenology group at Berkeley for stimulating discussions on the topic. 
We thank Nicola Bellomo, Kim Berghaus and Kim Boddy for discussions.
D.R.\ is supported by the NSF Grant No.\,PHYS-2014215, the DoE
HEP QuantISED award No.\,100495, and the Gordon and
Betty Moore Foundation Grant No.\,GBMF7946.
A.R.\ is supported by the
Boninchi Foundation for the project `PBHs in the Era of
GW Astronomy".

%%%%%%%%%%%%%%%%%%%%%%%%%%%%%%%%%%%%%%%%%%%%%%%%%%%%%%%%%%
%%%%%%%%%%%%%% REFERENCES %%%%%%%%%%%%%%%%%%%%%%%%%%%%%%%%

\end{document}